\documentclass[apl,twocolumn,showpacs,superscriptaddress,letterpaper]{revtex4-2}
\usepackage{graphicx,amsmath,amsfonts}
\usepackage{tabularx}
\usepackage{bm}
\usepackage{mathrsfs}
\usepackage{appendix}
\usepackage{bbm}
\usepackage{color}
\def \Ups {\Omega }
\def \vk {{\bf k}}
\def \vS {{\bf S}}
\def \vL {{\bf L}}
\def \mb {\mu_{\rm B}}
\def \vh {{\bf h}}
\def \vn {{\bf n}}
\def \vz {{\bf z}}
\def \vx {{\bf x}}
\def \vy {{\bf y}}
\def \vr {{\bf r}}
\def \vp {{\bf p}}
\def \xik {\xi_{\vk }}

\def \lz {{\hat{l}_z}}
\def \lk {{\hat{l}_{z\vk }}}

\def \bmu {{\boldsymbol{\mu}}}

\def \mB {\mu_{\rm B}}

\begin{document}

\title{{Orbital Angular Momentum of Magnons in Collinear Magnets}}

\author{Randy S. Fishman}
\email{corresponding author:  fishmanrs@ornl.gov
\newline
This manuscript has been authored in part by UT-Battelle, LLC, under contract DE-AC05-00OR22725 with the US Department of Energy (DOE). The US government retains and the publisher, by accepting the article for publication, acknowledges that the US government retains a nonexclusive, paid-up, irrevocable, worldwide license to publish or reproduce the published form of this manuscript, or allow others to do so, for US government purposes. DOE will provide public access to these results of federally sponsored research in accordance with the DOE Public Access Plan (http://energy.gov/downloads/doe-public-access-plan)}

\affiliation{Materials Science and Technology Division, Oak Ridge National Laboratory, Oak Ridge, Tennessee 37831, USA}
\author{Jason S. Gardner}
\affiliation{Materials Science and Technology Division, Oak Ridge National Laboratory, Oak Ridge, Tennessee 37831, USA}
\author{Satoshi Okamoto}
\affiliation{Materials Science and Technology Division, Oak Ridge National Laboratory, Oak Ridge, Tennessee 37831, USA}

\date{\today}

\begin{abstract}
	We study the orbital angular momentum of magnons for collinear ferromagnet (FM) and antiferromagnetic (AF) systems with
	nontrivial networks of exchange interactions.   The orbital angular momentum of magnons for AF and FM zig-zag 
	and honeycomb lattices becomes nonzero when the lattice contains two inequivalent sites and is largest at the avoided-crossing points or extremum of the frequency bands.
	Hence, the arrangement of exchange interactions may 
	play a more important role at producing the orbital angular momentum of magnons than the spin-orbit coupling energy and the resulting 
	non-collinear arrangement of spins.
		
	\end{abstract}

\keywords{spin-waves, orbital angular momentum}

\maketitle

For more than a century, scientists have been intrigued by the conversion of spin into orbital angular momentum (OAM) and vice versa.  
In 1915, A. Einstein and W.J. de Haas \cite{Einstein15} demonstrated that a change of magnetization can cause the container of that magnet to rotate.  
Also in 1915, S.J. Barnett \cite{Barnett15} demonstrated that the rotation of electrons can be converted into magnetization.  
In solids, the conversion of spin into orbital angular momentum is produced by the spin-orbit (SO) coupling.  
Recently, scientists have been searching for evidence of OAM \cite{pseudo, Streib21} in spin excitations, also known as magnons.  Whereas a magnon corresponding to
a single spin flip has spin ${\cal S} = \pm \hbar $, the OAM ${\cal L}$ of such a magnon is unknown.

Two main approaches have been employed to search for the OAM of magnons.  Because SO coupling is also responsible for
Dzyalloshinskii-Moriya (DM) interactions, Neumann {\it et al.} \cite{Neumann20} examined
the OAM of magnons associated with the non-collinear spin states produced by DM interactions.  
Other groups have investigated the OAM of magnons in confined geometries.  In a whispering gallery mode cavity, for example, circulating magnons with perpendicular OAM
can be excited on the surface of a FM sphere by incident light \cite{Haigh16, Sharma17, Osada18}.  Magnons with a range of orbital quantum numbers have been predicted for 
a FM nanocylinder that hosts a skyrmion at one end \cite{Jiang20}.
Quantum confinement of magnons 
has also been observed in a ferrite disk placed inside a microwave cavity \cite{Kamenetskii21}.
While approaches based on both SO coupling and confined geometries have achieved some success, 
they also require complex experiments and theories.
In an unrelated approach, Matsumoto and Murakami \cite{Mat11a} developed an expression for the OAM of FM magnons due to their ``self-rotation,"
which on average is opposed by the contribution of magnons to the edge current \cite{Mat11b, Li21}.

This Letter demonstrates that collinear magnets with tailored exchange geometries can generate magnons that exhibit OAM.
Results for both FM and AF zig-zag and honeycomb lattices in two dimensions indicate that 
the OAM becomes nonzero when the lattice contains two inequivalent sites and 
is greatest at the avoided-crossing points or extremum of the magnon bands.  
For FM zig-zag chains, the OAM vanishes when the upper and lower bands cross
but becomes quite large when the gap between the bands is small but nonzero. 
For FM honeycomb lattices, the upper
and lower bands carry opposite OAM when averaged over the Brillouin zone (BZ).  
For AF honeycomb lattices, the two degenerate magnon bands can be divided into major and minor branches that carry different OAM.
We shall see that the OAM and Berry curvature \cite{Xiao10} capture different aspects of the magnon band topology.

Formally, the classical equations of motion \cite{Tsukernik66, Garmatyuk68} for the dynamical magnetization ${\bmu}_i=2\mb \,\delta \vS_i$ at site $i$
produce 
the linear momentum ${\bf p}_i$ \cite{Landau60}:
\begin{equation}
p_{i\alpha }=\frac{1}{4\mb M_0} (\bmu_i \times \vn_i ) \cdot \frac{\partial \bmu_i}{\partial x_{\alpha }},
\end{equation}
where $M_0=2\mb S$ is the static magnetization for a spin ${\bf{S}}_i$ pointing along $\vn_i$ (a derivation of the classical OAM is provided in the Supplementary Material \cite{supm}).   
Using the $1/S$ quantization conditions
${\overline{\mu}_i}^+=\mu_{ix}n_{iz}+i\mu_{iy}=2\mb \sqrt{2S\hbar }\,a_i$ and ${\overline{\mu}_i}^-=\mu_{ix}n_{iz}-i\mu_{iy}=2\mb \sqrt{2S\hbar }\,a_i^{\dagger }$ for the dynamical magnetization
in terms of the local Boson operators $a_i$ and $a_i^{\dagger}$ satisfying the momentum-space
commutation relations $[a_{\vk }^{(r)}, a_{\vk'}^{(s)\dagger }]=\delta_{rs}\delta_{\vk,\vk'}$
and $[a_{\vk }^{(r)}, a_{\vk'}^{(s)}]=0$,
the quantized OAM along $\vz $ is given by
\begin{eqnarray}
{\cal L}_z &= &\sum_i (\vr_i \times \vp_i )\cdot \vz \nonumber \\
&=&\frac{\hbar}{2}\sum_{r=1}^M \sum_{\vk } \Bigl\{ a_{\vk}^{(r)}\,\lk \, a_{\vk }^{(r)\dagger } -a_{\vk }^{(r)\dagger }\, \lk \, a_{\vk }^{(r)}\Bigr\},
\end{eqnarray}
where $r$ and $s$ refer to the $M$ sites in the magnetic unit cell and 
\begin{equation}
\lk = -i \biggl( k_x\frac{\partial }{\partial k_y}-k_y\frac{\partial}{\partial k_x}\biggr)
\label{lk}
\end{equation}
is the OAM operator.
Transforming to the Boson operators $b_{\vk }^{(n)}$ and $b_{\vk }^{(n)\dagger }$
that diagonalize the Hamiltonian $H$, we define \cite{fishmanbook18}
\begin{eqnarray}
a_{\vk }^{(r)}&=&\sum_n \Bigl\{ X^{-1}(\vk )_{rn}\,b_{\vk}^{(n)} +X^{-1}(\vk )_{r,n+M}\,b_{-\vk}^{(n)\dagger }\Bigr\} , \\
a_{-\vk }^{(r)\dagger }&=&\sum_n \Bigl\{ X^{-1}(\vk )_{r+M,n}\,b_{\vk}^{(n)} +X^{-1}(\vk )_{r+M,n+M}\,b_{-\vk}^{(n)\dagger }\Bigr\} .
\nonumber
\end{eqnarray}
The zero-temperature expectation value of ${\cal L}_z$ 
for magnon state $b_{\vk }^{(n)\dagger }\vert 0\rangle =\vert \vk,n\rangle $ with frequency $\omega_n(\vk)$ is 
\begin{eqnarray}
{\cal L}_{zn}(\vk )&=&\langle \vk ,n \vert {\cal L}_z \vert \vk ,n \rangle \nonumber \\
&=& \frac{\hbar}{2}\sum_{r=1}^M\Bigl\{ 
X^{-1}(\vk )_{rn}\,\lk \, X^{-1}(\vk )_{rn}^* \nonumber \\
&&- X^{-1}(\vk)_{r+M,n}\,\lk \, X^{-1}(\vk )^*_{r+M,n}\Bigr\}.
\label{Lzdef}
\end{eqnarray}
For collinear spin states {\it without} DM interactions, $\underline{X}^{-1}(-\vk )=\underline{X}^{-1}(\vk )^*$ so that
${\cal L}_{zn }(\vk )=-{\cal L}_{zn }(-\vk )$ is an odd function of $\vk $. 

\begin{figure}
\begin{center}
\includegraphics[width=8cm]{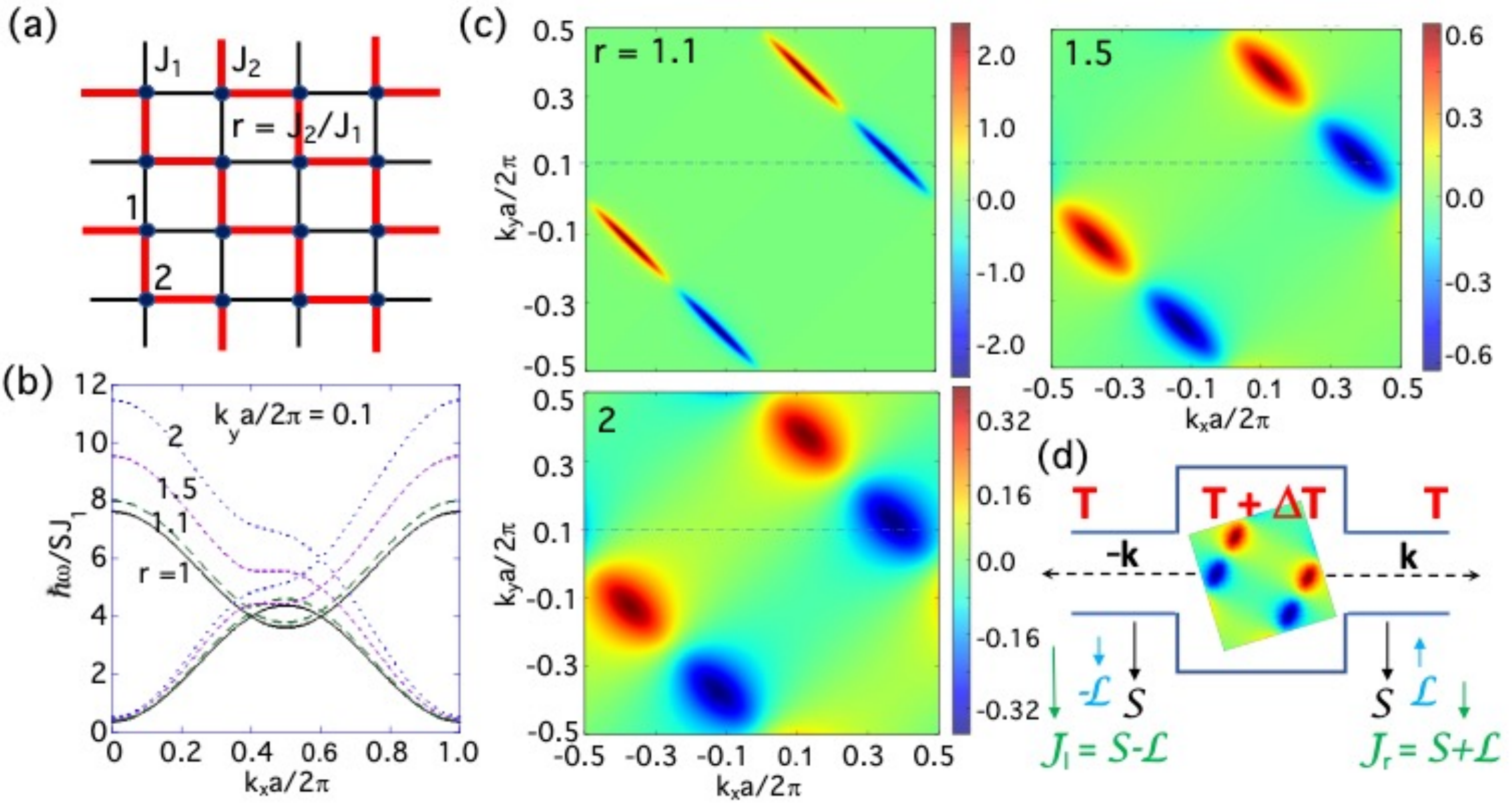}
\end{center}
\caption{{\it FM zig zag}.  (a) A square lattice with alternating FM exchange interactions $J_1$ and $J_2$.  (b) Magnon bands
for $k_ya/2\pi = 0.1$.  (c) The OAM ${\cal L}_{zn}(\vk )/\hbar $ graphed as a function of $\vk $ for different values of $r=J_2/J_1$.  
The dashed line shows $k_ya/2\pi = 0.1$  (d)  Magnons of an $r=2$ FM zig-zag material traveling with opposite momenta $\pm \vk $ and OAM $\pm {\cal L}$ 
in a temperature gradient.}
\label{Fig1}
\end{figure}

{\it i. FM zig zag}.  Our first case study is the square lattice shown in Fig.\,1(a) with alternating FM bonds $J_1>0$ and $J_2>0$
coupling sites 1 and 2 with spins up.  Second order in the operator ${\bf v}_{\vk } =(a_{\vk }^{(1)},a_{\vk }^{(2)},a_{-\vk}^{(1)\dagger },a_{-\vk }^{(2)\dagger })$,
the Hamiltonian $H_2=\sum_\vk {\bf v}_{\vk}^{\dagger }\cdot \underline{{\it L}}(\vk )\cdot {\bf v}_{\vk }$ is defined in terms of the matrix
\begin{equation}
\underline{L}(\vk ) =
(J_1+J_2)S
\left(
\begin{array}{cccc}
1 & -\Psi_{\vk}^* & 0 & 0 \\
-\Psi_{\vk } & 1 & 0 & 0 \\
0 & 0 & 1 & - \Psi_{\vk }^*\\
0 & 0 &-\Psi_{\vk } & 1\\
\end{array} \right),
\end{equation}
where $\Psi_{\vk }=\bigl(J_1\xik^*+J_2\xik \bigr)/2(J_1+J_2)$ with
$\xik =\exp(ik_xa) + \exp(ik_ya)$.
To study the magnon dynamics, we must diagonalize $\underline{L}\cdot \underline{N}$, where
\begin{equation}
\underline{N} =
\left(
\begin{array}{cc}
\underline{I} &0 \\
0 & -\underline{I} \\
\end{array} \right)
\end{equation}
and $\underline{I}$ is the two-dimensional identity matrix.
Using the relation $\underline{N}\cdot \underline{X}^{\dagger }(\vk )\cdot \underline{N}=\underline{X}^{-1}(\vk)$ to normalize the 
eigenvectors \cite{fishmanbook18} $X^{-1}(\vk )_{rn}$, we find
\begin{equation}
\underline{X}^{-1}(\vk ) =
\frac{1}{\sqrt{2}\,\Psi_{\vk }^*}
\left(
\begin{array}{cccc}
-\Psi_{\vk}^* & \Psi_{\vk }^* & 0 & 0 \\
\vert \Psi_{\vk }\vert  & \vert \Psi_{\vk }\vert & 0 & 0 \\
0 & 0 & -\Psi_{\vk }^* & \Psi_{\vk }^*\\
0 & 0 &\vert \Psi_{\vk }\vert  & \vert \Psi_{\vk }\vert  \\
\end{array} \right).
\label{xin1}
\end{equation}
It is then simple to show that 
\begin{equation}
\label{lz1}
{\cal L}_{zn}(\vk )=\frac{\hbar}{4}\frac{\Psi_{\vk }}{\vert \Psi_{\vk }\vert }\,\lk \,
\frac{\Psi_{\vk }^*}{\vert \Psi_{\vk }\vert }
\end{equation}
is the same for magnon bands $n=1$ and 2 with energies $\hbar \omega_{1,2}(\vk )=(J_1+J_2)S(1\pm \vert \Psi_{\vk }\vert )$.

Results for ${\cal L}_{zn}(\vk )/\hbar $ are plotted as a function of $r=J_2/J_1$ in Fig.\,1(c) \cite{kper}.  Not surprisingly, ${\cal L}_{zn}(\vk )$ vanishes for a square-lattice FM with
$r=1$.  Comparing the ``hot spots" in Fig.\,1(c) for $r=1.1$ with the magnon bands in Fig.\,1(b) for $k_ya/2\pi =0.1$, we see that the OAM is largest ($\sim 2\hbar $) at the 
avoided-crossing points $\vk^*$ of bands 1 and 2 near $k_xa/2\pi  = 0.4$.   As $r$ increases, the gap between the bands grows, the 
region of large $\vert {\cal L}_{zn}(\vk )\vert $ spreads out in $\vk $ space, 
and its amplitude decreases.  For very large $r$, the regions of large positive and negative ${\cal L}_{zn}(\vk )$ stretch into stripes.
The wavevectors $\vk^*$ are associated with a sign change in the Berry curvature \cite{Xiao10, supm}.

\begin{figure}
\begin{center}
\includegraphics[width=8 cm]{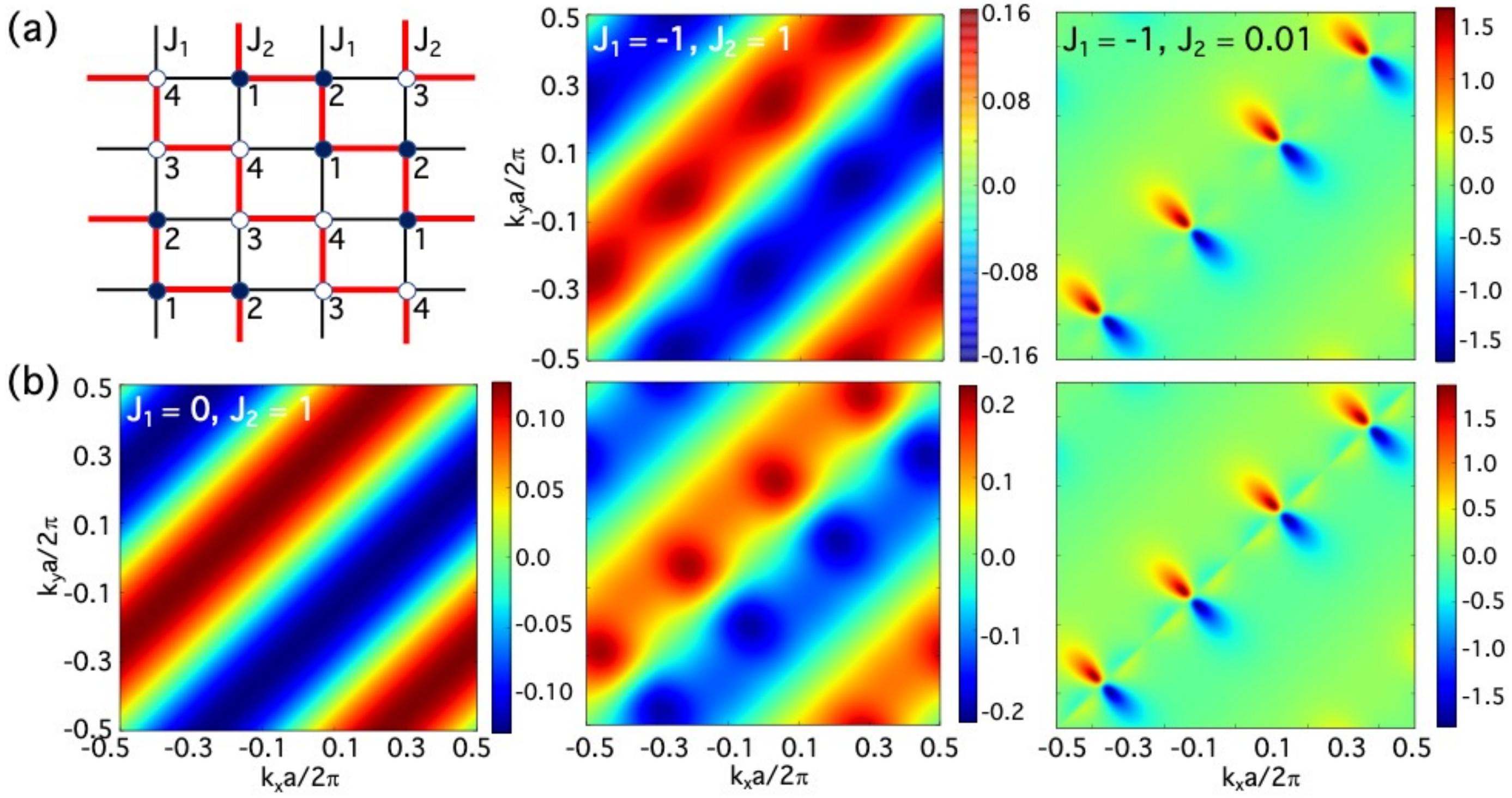}
\end{center}
\caption{{\it AF zig zag}.  (a) A square lattice with FM exchange $J_2 >0$ on zig-zag chains with up (closed circles) or down (open circles) spins
coupled by AF exchange $J_1 < 0$.  (b) The OAM for upper (top) and lower (bottom) bands versus $\vk $ for different values of 
$J_1$ and $J_2$.}
\label{Fig2}
\end{figure}

{\it ii. AF zig zag}.  For the square lattice in Fig.\,2(a), we take $J_1 < 0$  and $J_2>0$ so that sites $1$ and $2$ have spins up
while sites $3$ and $4$ have spins down.   Although $\underline{L}(\vk )$ is 8 dimensional, it breaks into the
two identical $4\times 4$ matrices
\begin{equation}
\underline{L}(\vk )' =
(J_2-J_1)S
\left(
\begin{array}{cccc}
1 & -\gamma_2\xik & 0 &  \gamma_1\xik^* \\
-\gamma_2 \xik^* & 1 & \gamma_1\xik  & 0 \\
0 & \gamma_1\xik^*  & 1 & - \gamma_2\xik \\
\gamma_1\xik  & 0 &-\gamma_2\xik^*  & 1\\
\end{array} \right)
\end{equation}
with doubly degenerate magnon energies
\begin{eqnarray}
\hbar \omega_{1,2}(\vk )&=& 2(J_2-J_1)S\biggl\{ 1 - (\gamma_1^2-\gamma_2^2)\vert \xik\vert^2 \nonumber \\
&\pm &\gamma_2 \sqrt{ \gamma_1^2 (\xik^2 -\xik^{*2})^2+4\vert \xik\vert^2} \biggr\}^{1/2},
\end{eqnarray}
where $\gamma_n=J_n/2(J_2-J_1)$.

While no simple analytic expression for the 
OAM is possible, we readily obtain the numerical solutions in Fig.\,2(b).  For $J_1=0$, the zig-zag chains are isolated
from each another and the numerical solution is identical to one for FM zig-zag chains.  Hence, the two bands
have the same OAM.  When $J_1=-J_2$, the lower band exhibits a larger amplitude of the OAM than the upper band, as seen in the central panel of Fig.\,2(b).
When $J_2=0.01$ and $J_1=-1$, the FM interaction within each zig-zag chain is very weak while the AF
interaction between chains is strong.  Then the OAM is only significant around discrete points $\vk^*$ along the line $k_x = k_y$. 
As expected, the OAM vanishes as $J_2/\vert J_1\vert \rightarrow 0$.

\begin{figure}
\begin{center}
\includegraphics[width=8cm]{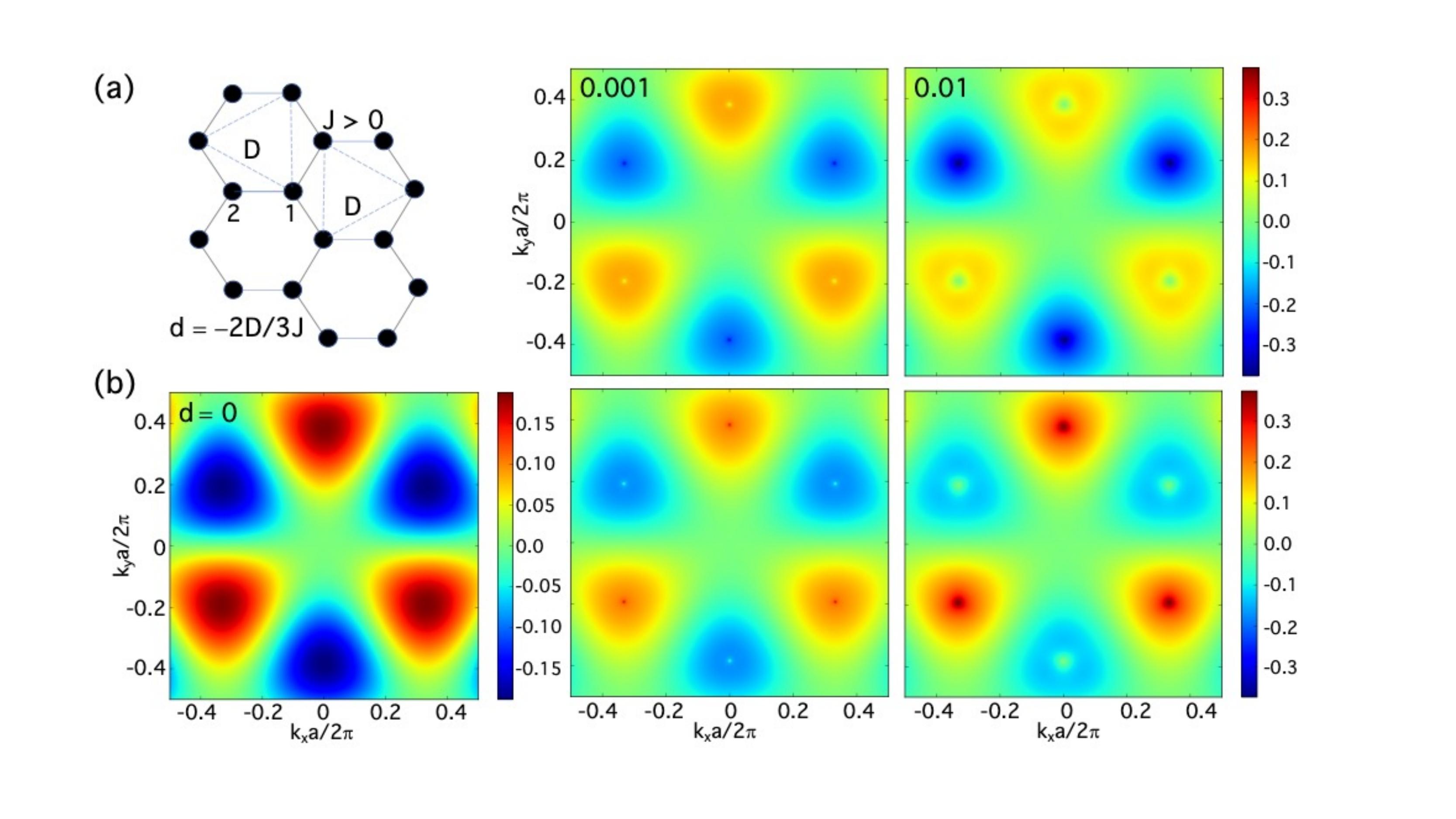}
\end{center}
\caption{{\it FM honeycomb}.  (a) A honeycomb lattice with FM exchange $J>0$ and DM interaction $D$ between next-nearest
neighbors.  (b) The OAM for the upper (top) and lower (bottom) bands versus $\vk $ for different values of $d=-2D/3J$.}
\label{Fig3}
\end{figure}

{\it iii. FM honeycomb}.  We now consider 
the honeycomb lattice shown in Fig.\,3(a) with FM exchange coupling $J>0$.
Provided that the easy-axis anisotropy $-K\sum_i S_{iz}^2$ is sufficiently strong, 
we may also add a DM interaction $D$ between next-neighbor sites without tilting the spins.
We then find 
\begin{equation}
\underline{L}(\vk ) =
\frac{3JS}{2} \left(
\begin{array}{cccc}
1-G_{\vk }  & -\Gamma_{\vk}^* & 0 & 0 \\
-\Gamma_{\vk } & 1+G_{\vk }  & 0 & 0 \\
0 & 0 & 1+G_{\vk}  & - \Gamma_{\vk }^*\\
0 & 0 &-\Gamma_{\vk } & 1-G_{\vk} \\
\end{array} \right),
\end{equation}
where $G_{\vk }=d \,\Theta_{\vk }$ with $d = -2D/3J$,
$\Theta_{\vk} = 4\cos(3k_xa/2)$ $\sin(\sqrt{3} k_ya/2)-2\sin(\sqrt{3}k_ya)$, and 
\begin{equation}
\Gamma_{\vk} =\frac{1}{3}\Bigl\{ e^{ik_xa}+e^{-i(k_x+\sqrt{3}k_y)a/2}
+e^{-i(k_x-\sqrt{3}k_y)a/2}\Bigr\}.
\end{equation}
Because the anisotropy $\kappa =K/J$ merely shifts the magnon energies 
$\hbar \omega_{1,2}(\vk)=3JS ( 1 - \kappa \pm g_{\vk } )$ 
with $g_{\vk }=\sqrt{\vert \Gamma_{\vk }\vert^2 +G_{\vk }^2}$
but does not affect the OAM,
we neglect its contribution to $\underline{L}(\vk )$.  
After the usual manipulations, we find $
X^{-1}(\vk)_{11}=-1/2c_1g_{\vk }$, $X^{-1}(\vk)_{12}=1/2c_2g_{\vk },$
$X^{-1}(\vk)_{21}=(G_{\vk}+g_{\vk })/2c_1\Gamma_{\vk }^* g_{\vk }$,
and 
$X^{-1}(\vk)_{22}=-(G_{\vk}-g_{\vk })/2c_2\Gamma_{\vk }^* g_{\vk }$,
where
$1/\vert c_1\vert^2 =2g_{\vk }(g_{\vk }-G_{\vk })$ and
$1/\vert c_2\vert^2 =2 g_{\vk } (g_{\vk }+G_{\vk })$.
The 31, 32, 41, and 42 matrix elements of $\underline{X}^{-1}(\vk)$ vanish.

For $d=0$, the upper and lower band frequencies $\omega_1(\vk )$ and $\omega_2(\vk )$ cross at $\vk^*=(1/3,\sqrt{3}/9)(2\pi /a)$ and equivalent points throughout the BZ.  
With 
\begin {equation}
{\cal L}_{zn}(\vk )=\frac{\hbar }{4}\frac{\Gamma_{\vk}}{ \vert \Gamma_{\vk }\vert }\,
 \lk \, \frac{\Gamma_{\vk }^*}{\vert \Gamma_{\vk }\vert },
 \label{lz3}
 \end{equation}
the OAM is the same for both bands.  Notice that this expression is the same as Eq.\,(\ref{lz1}) for ${\cal L}_{zn}(\vk )$ of the FM zig-zag lattice with $\Psi_{\vk }$ replaced by $\Gamma_{\vk }$. 
As seen in Fig.\,3(b), ${\cal L}_{zn}(\vk )/\hbar $
has modest values of $\pm 3/16 = \pm 0.1875$ at $\vk^*$ \cite{kper, fishun}.

Since DM interactions change sign upon spatial inversion,
${\cal L}_{zn}(\vk )/\hbar $ contains 
both even and odd terms with respect to $\vk $ due to the $G_{\vk }=-G_{-\vk} \sim d$ functions in $\underline{X}^{-1}(\vk )$.
For $d>0$, the averages of ${\cal L}_{z1}(\vk )/\hbar $ and ${\cal L}_{z2}(\vk )/\hbar $ over the BZ are negative and positive, respectively.
With increasing $d$, a gap opens between the two magnon bands and $\vert {\cal L}_{zn}(\vk )\vert $ grows at the avoided-crossings points $\vk^*$.  
For $d=0.01$, the largest values of the OAM at $\vk^*$ are about $\pm 0.38\hbar $.
The Berry curvature \cite{Xiao10} of the FM honeycomb lattice is discussed in the Supplementary Material \cite{supm}.

\begin{figure}
\begin{center}
\includegraphics[width=8cm]{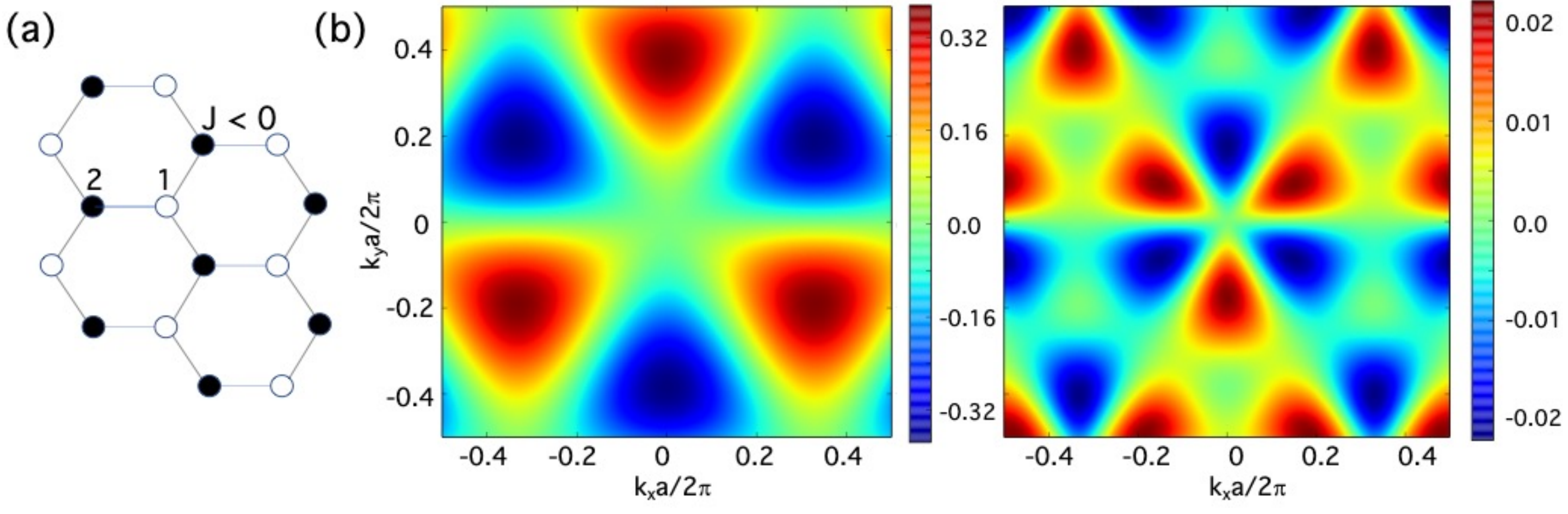}
\end{center}
\caption{{\it AF honeycomb}.  (a) A honeycomb lattice with AF exchange $J<0$ between up (closed circles, site 2) and down (open circles, site 1) spins.
(b) The OAM of the major (left) and minor (right) bands versus $\vk $ for anisotropy $\kappa =0$.}
\label{Fig4}
\end{figure}

{\it iv. AF honeycomb}.  The final case study is the honeycomb lattice sketched in Fig.\,4(a) with AF exchange  
$J<0$ between alternating up and down spins.  Since it shifts the magnon energies but 
does not affect the OAM, the DM interaction is neglected in the following discussion.  We obtain
\begin{equation}
\underline{L}(\vk ) =
-\frac{3JS}{2} \left(
\begin{array}{cccc}
1+\kappa  & 0 & 0 & -\Gamma_{\vk}^* \\
0 & 1+\kappa   &  -\Gamma_{\vk} & 0\\
0 & -\Gamma_{\vk}^* & 1+\kappa  & 0\\
-\Gamma_{\vk} & 0 &0 & 1+\kappa  \\
\end{array} \right).
\end{equation}
The usual procedure yields
$X^{-1}(\vk )_{11}=-1/2c_1f_{\vk },$
$X^{-1}(\vk )_{32}= 1/2c_2f_{\vk },$
$X^{-1}(\vk )_{22}=(f_{\vk }+1+\kappa )/2c_2\Gamma_{\vk }^*f_{\vk }$,
and 
$X^{-1}(\vk )_{41}=(f_{\vk }-1-\kappa )/2c_1\Gamma_{\vk }^*f_{\vk },$
where 
$1/\vert c_1\vert^2= 2f_{\vk }(1+\kappa +f_{\vk })$
and
$1/\vert c_2\vert^2= 2f_{\vk }(1+\kappa -f_{\vk })$
with $f_{\vk } =\sqrt{(1+\kappa )^2-\vert \Gamma_{\vk }\vert^2}$.
Other matrix elements of $X^{-1}(\vk )_{rn} $ for modes $n=1$ and 2 vanish.

Surprisingly, the doubly degenerate magnon bands with 
energies $\hbar \omega_{1,2}(\vk )=3\vert J\vert S\sqrt{(1+\kappa )^2-\vert \Gamma_{\vk }\vert^2}$ 
exhibit distinct OAMs with 
\begin {eqnarray}
{\cal L}_{z1}(\vk )&=&\frac{\hbar }{4}\frac{1+\kappa +f_{\vk }}{f_{\vk }} \,
\frac{\Gamma_{\vk} }{\vert \Gamma_{\vk }\vert }\,\lk \, \frac{\Gamma_{\vk }^* }{\vert \Gamma_{\vk }\vert },\\
{\cal L}_{z2}(\vk )&=&-\frac{\hbar }{4}\frac{1+\kappa -f_{\vk }}{f_{\vk }}\,
\frac{\Gamma_{\vk} }{\vert \Gamma_{\vk }\vert }\,\lk \, \frac{\Gamma_{\vk }^* }{\vert \Gamma_{\vk }\vert },
\end{eqnarray}
and ratio
${\cal L}_{z1}(\vk )/{\cal L}_{z2}(\vk ) = -(1+\kappa +f_{\vk })/(1+\kappa -f_{\vk }) < 0$.
As seen in Fig.\,4(b) for $\kappa =0$, the major and minor bands have different patterns for ${\cal L}_{zn}(\vk )$
but are both threefold symmetric.  
The maxima in $\vert {\cal L}_{z1}(\vk )\vert/\hbar $ of $3/8$ \cite{fishun} appear at 
points $\vk^*$ where $\Gamma_{\vk }$ vanishes and $\hbar \omega_n(\vk )$ reaches a maximum of $3\vert J\vert S$.  
Those points coincide with the avoided-crossing points $\vk^*$ of the non-degenerate bands for the FM honeycomb lattice. 

For $\kappa >0$, the average OAM
${\cal L}_{{\rm av }}(\vk )= ({\cal L}_{z1}(\vk )+{\cal L}_{z2}(\vk ))/2$ of the major and minor
bands of the AF honeycomb lattice equals the OAM of the $d=0$ FM honeycomb lattice given by Eq.\,(\ref{lz3}) and plotted in Fig.\,3.
We emphasize that the major and minor bands of the AF honeycomb lattice are identical in every other respect.  For example, their 
spin-spin correlation functions $S_{\alpha \beta}(\vk ,\omega )$ are equal \cite{supm}.  

The topological nature of quasiparticles in solids is often characterized by their Berry phase \cite{Xiao10}. 
In momentum space, the Berry curvature is given by
\begin{eqnarray}
\Omega_n (\vk)=\frac{i}{2 \pi} 
\Bigl\{ {\bm \nabla}_{\vk} \times
\langle u_n(\vk ) | {\bm \nabla}_{\vk} u_n(\vk) \rangle \Bigr\}\cdot \vz, 
\label{EqBerry}
\end{eqnarray}
where $|u_n(\vk )\rangle$ is the single-particle wave function of band $n$ and $\langle u_n(\vk )| {\bm \nabla}_{\vk} u_n(\vk ) \rangle$ is called the Berry connection. 
Integrating $\Omega_n (\vk)$ over the BZ then gives the Chern number ${\cal C}_n$. 
The connection between the Berry curvature and the OAM is clarified by rewriting Eq.\,(\ref{Lzdef}) as 
\begin{equation}
\label{aex}
{\cal L}_{z n} (\vk) =-\frac{i \hbar}{2} \Bigl\{\vk \times \langle u_n(\vk )| {\bm \nabla}_{\vk} u_n(\vk ) \rangle \Bigr\}\cdot \vz. 
\end{equation}
Thus, while the Berry curvature is the curl of the Berry connection, the OAM is the cross product of the momentum $\vk$ and Berry connection. 

At low energies and momenta, Eq.\,(\ref{aex}) reduces to the expression of Matsumoto and Murakami \cite{Mat11a, Mat11b} for FM magnons, which 
was parameterized in terms of an effective mass $m^*$.
Since we are interested in the OAM of both FM and AF magnons throughout the BZ, we prefer using the more general expression given above.
Because it is produced by SOC, the OAM discussed in Ref.\,\cite{Neumann20} is not related to the one described by Eq.\,(\ref{aex}).

Theoretically, the OAM predicted in this paper vanishes for mode $n$ if the matrix elements $X^{-1}(\vk )_{rn}$ and $X^{-1}(\vk )_{r+M,n}$ can be simultaneously rotated 
onto the real axis by a suitable choice of normalization factor $c_n$ \cite{supm}.
This generates terms like $\exp(ik_xa)$
and $\exp(ik_ya)$ that are not mixed with their complex conjugates in $\underline{L}(\vk )$, $X^{-1}(\vk )_{rn}$, and $X^{-1}(\vk )_{r+M,n}$.

Whenever magnons exhibit OAM, the lattice contains two inequivalent sites either due to exchange (cases $i$ and $ii$) or structure (cases $iii$ and $iv$).
In such a non-Bravais lattice, the violation of inversion symmetry about each site creates preferred channels for the magnons and an asymmetry
in $\vk $ space that produces the OAM.
In that sense, the present work follows in the spirit of earlier work on magnon confinement in spherical 
\cite{Haigh16, Sharma17, Osada18} and cylindrical \cite{Jiang20, Kamenetskii21} geometries.
We surmise that it may be easier to generate and control the OAM of magnons by designing devices with tailored exchange interactions than with 
customized SO couplings and spin textures.

In all four case studies, the largest OAM appears at the crossing points or extremum $\vk^*$ of the magnon bands.
For the FM zig-zag lattice, a slight increase of $r=J_2/J_1$ from 1 has  
a huge effect on the OAM because it creates two inequivalent magnetic sites while opening a gap between the magnon bands at $\vk^*$.  
Increasing $r>1$ further reduces the OAM while widening the gap between the magnon bands.
Since the FM honeycomb lattice with $D=0$ already contains two inequivalent sites, its magnons exhibit nonzero OAM at wavevectors $\vk^*$ and 
elsewhere throughout the BZ.  By breaking the odd symmetry of ${\cal L}_{zn}(\vk )$, a nonzero $D$ allows the upper and lower magnon bands to
carry a net OAM when averaged over the BZ.  Consequently, larger values of the OAM appear at $\vk^*$.
Because it breaks the degeneracy of otherwise identical bands, the OAM of an AF honeycomb lattice is particularly intriguing.

This work opens the gateway for the future experimental study of the OAM of magnons in collinear spin systems.  
While bulk zig-zag systems with $J_1\approx J_2>0$ (case $i$) are difficult to experimentally identify
due to their similar exchange constants, many experimental systems can be described as zig zags 
coupled by AF exchange $J_1<0$ (case $ii$). 
AF-coupled zig-zag chains decorate the quasi-two-dimensional honeycomb lattice compound 
Na$_2$Co$_2$TeO$_6$ \cite{Bera17}, the transition-metal thiophosphates XPS$_3$ (X = Fe or Ni) \cite{Wildes15, lan16, Zhang21},
and iridium-based compounds like Na$_2$IrO$_3$ \cite{Ye12}. 
Both the honeycomb sublattice of Li$_3$Ni$_2$SbO$_6$ \cite{Kurbakov17} and
the square AF sublattice of Ba$_2$Mn(PO$_4$)$_2$ \cite{Yogi19} also contain zig-zag chains.
While many Ruddlesden-Popper manganites have zig-zag chains with AF correlations \cite{Salamon01},  
the metallic manganite La$_{0.67}$Ca$_{0.33}$MnO$_3$ has zig-zag chains running within square AF
{\it{ab}}-planes \cite{Panopoulos18}.
Due to their photoluminescent properties, many of these materials 
are candidates for opto-spintronics, which provides avenues to probe or perturb the 
OAM of magnons.   

The magnetic phase diagrams of honeycomb systems with 
chemical formula ABX$_3$ were reviewed by Sivadas {\it et al.} \cite{Sivadas15}.
Examples of FM honeycomb lattices (case $iii$) are CrSiTe$_3$ and CrGeTe$_3$ \cite{Carteaux95a, Carteaux95b, Casto15}.  Another well-known 
Cr-based FM honeycomb system is CrI$_3$ \cite{McGuire15}, which has topological magnon excitations that were studied by Chen {\it et al.} \cite{Chen18t}.
AF honeycomb lattices (case $iv$) are found in MnPS$_3$ and MnPSe$_3$ \cite{Wildes98}.  
 

There are many physical consequences connected with the predicted OAM of magnons, including its effect on magnon decay rates, the 
scattering by photons and phonons, and the scattering of magnons in thermal gradients (see Fig.\, 1(d)) \cite{Cheng16}. 
Once a magnon with momentum $\vk $ and OAM ${\cal L}_{zn}(\vk)$ is created, conservation of total angular momentum ${\cal J} = {\cal S}+{\cal L}$ (spin plus orbital)
due to dipolar interactions has been demonstrated for small $ka$ \cite{Tsukernik66, Garmatyuk68} even in the absence of SO coupling.  
Most present measurements of magnon transport do not probe the OAM ${\cal L}_{zn}(\vk )$,
which averages to zero over magnon bands within the BZ.  While many issues remain to be explored, including the generalization of this 
work for non-collinear spin states, we have established that the 
magnons of two-dimensional collinear magnets can 
carry significant OAM provided that the exchange interactions meet some easily satisfied conditions.  We hope that
future theoretical and experimental work will explore the nature of that OAM and how it can be used to understand and control
the properties of magnons in magnetic materials.


We acknowledge useful conversations with D. Xiao and R. deSousa.
Research sponsored by the Laboratory Directors Fund of Oak Ridge National Laboratory.  
The data that support the findings of this study are available from the corresponding author
upon reasonable request.


\onecolumngrid
\newpage
\begin{center}
{\large \bf Supplementary Material:  Orbital Angular Momentum of Magnons in Collinear Spin Systems}\\
\vspace{1em}
Randy S. Fishman, Jason S. Gardner, and Satoshi Okamoto$^1$ \\
\vspace{0.5em}
{\small \it $^1$Materials Science and Technology Division, Oak Ridge National Laboratory, Oak Ridge, Tennessee 37831, USA}\\
\end{center}

\def \Ups {\Omega }
\def \vk {{\bf k}}
\def \vS {{\bf S}}
\def \vL {{\bf L}}
\def \mb {\mu_{\rm B}}
\def \vh {{\bf h}}
\def \vn {{\bf n}}
\def \vz {{\bf z}}
\def \vx {{\bf x}}
\def \vy {{\bf y}}
\def \vr {{\bf r}}
\def \vp {{\bf p}}
\def \vq {{\bf q}}
\def \lz {{\hat{l}_z}}
\def \lq {{\hat{l}_z(\vq)}}
\def \lk {{\hat{l}_{z\vk}}}
\def \re {{\rm Re}\,}
\def \im {{\rm Im}\,}
\def \kov {\overline{k}}

\def \bmu {{\boldsymbol{\mu}}}

\def \mB {\mu_{\rm B}}

\setcounter{equation}{0}
\setcounter{figure}{0}
\setcounter{table}{0}
\setcounter{page}{1}
\makeatletter
\renewcommand{\theequation}{S\arabic{equation}}
\renewcommand{\thefigure}{S\arabic{figure}}
\renewcommand{\bibnumfmt}[1]{[S#1]}
\renewcommand{\citenumfont}[1]{S#1}

\twocolumngrid

\section*{Classical equations of motion, linear momentum, and OAM}

We briefly review the classical equations of motion and Lagrangian formulation originally presented in Refs.\,\cite{Tsukernik66, Garmatyuk68}
for collinear spins.
A general Hamiltonian can be written in terms of the magnetization ${\bf M}_i$ as
\begin{eqnarray}
H&=& -\sum_{i=1}^M J_{\alpha \beta } \, M_{i\alpha }M_{i+1, \beta } -\sum_{i=1}^M {\bf M}_i \cdot {\bf h}_i \nonumber \\
&-&\frac{\kappa }{2}\sum_{i=1}^{M} M_{iz}^2
- \frac{1}{8\pi }\sum_{i=1}^M h_i^2,
\end{eqnarray}
where we only consider one dimension for simplicity and the magnetic unit cell contains $M$ sites.  Unless explicitly indicated, repeated Greek indices are summed. 
The magnetostatics equations for the magnetic dipole field ${\bf h}_i$ are
$\nabla \times {\bf h}_i=0$ and $\nabla \cdot {\bf h}_i=-4\pi \nabla \cdot {\bf M}_i$.  The first can be satisfied by defining a scalar potential
$\phi_i$ as ${\bf h}_i = \nabla \phi_i$.  Asymmetric exchange interactions like the DM interaction may be included in $J_{\alpha \beta }$.
It is also easy to include further-neighbor exchange interactions.
The total magnetization ${\bf M}_i$ can be written in terms of the dynamical magnetization $\boldsymbol{\mu}_i$ as
\begin{equation}
{\bf{M}}_i=\boldsymbol{\mu }_i +\vn_i \sqrt{M_0^2 - \mu_i^2},
\end{equation}
where the static magnetization ${\bf{M}}_{0i} = g S\,\vn_i $ ($g=2\mb $) lies along $\vn_i$ and $\boldsymbol{\mu}_i\cdot \vn_i =0$. 
Defining the effective field
\begin{equation}
{\bf H}_{{\rm eff}, i} = -\frac{\partial E }{\partial {\bf M}_i}
\end{equation}
in terms of the energy $E=\langle H\rangle $, the equations of motion for 
$\boldsymbol{\mu}_i$ are obtained by 
expanding
\begin{equation}
\frac{\partial {\bf M}_i}{\partial t}=-g \Bigl({\bf M}_i \times {\bf H}_{{\rm eff},i}\Bigr)
\end{equation}
to first order in $\boldsymbol{\mu }_i$:
\begin{eqnarray}
\dot{\boldsymbol{\mu}_i}&=& -g \Bigl\{ {\bf M}_{0i} \times \Bigl[ {\bf h}_i 
+\underline{J}\cdot \bigl(\boldsymbol{\mu}_{i+1}+\boldsymbol{\mu}_{i-1}\bigr) \nonumber \\
&-&\kappa \,\boldsymbol{\mu}_i \Bigr] \Bigr\},
\end{eqnarray}
which assumes that $\vn_i = \pm \vz $ for each site in the unit cell, i.e. a collinear spin state.

Alternatively, we may directly expand 
$H$ in powers of $\boldsymbol{\mu}_i$ to obtain $H=E+H_2+\ldots $ where
\begin{eqnarray}
H_2&=& -\sum_{i=1}^M J_{\alpha \beta } \, \mu_{i\alpha }\mu_{i+1, \beta } -\sum_{i=1}^M \boldsymbol{\mu}_i \cdot {\bf h}_i
\nonumber \\
&-&\frac{\kappa }{2}\sum_{i=1}^{M} \mu_{iz}^2
- \frac{1}{8\pi }\sum_{i=1}^M h_i^2.
\end{eqnarray}
If the Lagrangian is written \cite{Landau60} in terms of $\boldsymbol{\mu}_i$, $\dot{\boldsymbol{\mu}}_i$, and $\phi_i$ as
\begin{equation}
L=\frac{1}{2g M_0}\sum_{i=1}^M \bigl( \dot{\boldsymbol{\mu}}_i\times \vn_i \bigr) \cdot \boldsymbol{\mu}_i - H_2,
\end{equation}
then the Hamiltonian equations of motion for $\boldsymbol{\mu}_i$ given above can also be obtained from the Euler-Lagrange equations
\begin{equation}
\frac{d}{dt}\frac{\partial L}{\partial \dot{\mu}_{i\alpha }}=\frac{\partial L}{\partial \mu_{i\alpha }}.
\end{equation}
Based on the Lagrangian $L$, the energy-momentum tensor \cite{Landau60} $T_{\alpha \beta }$ is given by 
\begin{eqnarray}
T_{\alpha \beta } &=& L \,\delta_{\alpha \beta } - \sum_i  \frac{ \partial \boldsymbol{\mu}_i }{\partial x_{\alpha }}\cdot \frac{\partial L}{\partial (\partial \boldsymbol{\mu}_i/\partial x_{\beta })}
\nonumber \\
&-& \sum_i \frac{\partial \phi_i}{\partial x_{\alpha }}\frac{\partial L}{\partial (\partial \phi_i /\partial x_{\beta })},
\end{eqnarray}
where $x_{\alpha }=x,y,z$, or $t$ for $\alpha =1,2,3$, or 4, respectively.
It follows that the momentum density ($\alpha \ne 4,\, \beta = 4$) is
\begin{eqnarray}
p_{\alpha } =T_{\alpha 4} &=& -\sum_i \frac{\partial \boldsymbol{\mu}_i }{\partial x_{\alpha }}\cdot \frac{\partial L}{\partial \dot{\boldsymbol{\mu}_i}}\nonumber \\
&=& - \frac{1}{2g M_0} \sum_i \bigl( \vn_i \times \boldsymbol{\mu }_i\bigr)\cdot \frac{\partial \boldsymbol{\mu}_i}{\partial x_{\alpha }}.
\end{eqnarray}
Writing $p_{\alpha }=\sum_i p_{i\alpha }$ gives the expression for $p_{i\alpha }$ stated in Eq.\,(1) of the main paper.
The momentum $p_{\alpha}$ then satisfies the continuity relation
\begin{equation}
\frac{dp_{\alpha }}{dt} +\sum_{\beta =1}^{3} \frac{\partial T_{\alpha \beta }}{\partial x_{\beta }}=0
\end{equation}
for $\alpha =1,2,3$.

Transforming to the local reference frame of the spin (in the same spirit as in spin-wave theory \cite{fishmanbook18}), we use the local spin 
variables $\overline{\boldsymbol{\mu}}_i$ given by
\begin{eqnarray}
\overline{\mu}_{ix}&=&n_{iz} \, \mu_{ix},\\
\overline{\mu}_{iy}&=&\mu_{iy},\\
\overline{\mu}_{iz}&=&n_{iz} \,\mu_{iz}, 
\end{eqnarray}
with ${\overline\mu_i}^{\pm }=\overline{\mu}_{ix}\pm i \overline{\mu}_{iy}$
to find the total OAM
\begin{eqnarray}
{\cal L}_z &=&\sum_{i=1}^N ({\bf r_i}\times {\bf p}_i )_z \nonumber \\
&=& \frac{1}{4gM_0}\sum_{i=1}^N \Bigl\{ {\overline{\mu}_i}^+  \,\hat{l}_{zi} \, {\overline{\mu}_i}^- -{\overline{\mu}_i}^- \, \hat{l}_{zi} \, {\overline{\mu}_i}^+ \Bigr\},
\end{eqnarray}
where
\begin{equation}
\hat{l}_{zi} = -i \bigg( x_i \frac{\partial}{\partial y_i} - y_i\frac{\partial }{\partial x_i}\bigg)
\end{equation}
is the classical OAM operator in real space.

\section*{Symmetry Relations and the FM zig-zag lattice}

In order to clarify 
the symmetry relations for $\underline{X}^{-1}(\vk )$, we review some details of the OAM solution for the FM zig-zag lattice
(case $i$).
Based on the $\underline{L}(\vk )$ matrix in Eq.\,(6) of the main paper, 
we find that the eigenvectors of $\underline{L}(\vk )\cdot \underline{N}$ are
\begin{eqnarray}
X(\vk)_{1j}^*&=&c_1^*(-\vert \Psi_{\vk }\vert , \Psi_{\vk },0,0),\\
X(\vk)_{2j}^*&=&c_2^*(\vert \Psi_{\vk }\vert , \Psi_{\vk },0,0),\\
X(\vk)_{3j}^*&=&c_3(0,0,-\vert \Psi_{\vk }\vert , \Psi_{\vk }),\\
X(\vk)_{4j}^*&=&c_4(0,0,\vert \Psi_{\vk }\vert , \Psi_{\vk }),
\end{eqnarray}
where
\begin{equation}
\Psi_{\vk }=\frac{J_1(e^{-ik_xa}+e^{-ik_ya}) +J_2 (e^{ik_xa}+e^{ik_ya})  }{2(J_1+J_2)}.
\end{equation}
Hence,
\begin{equation}
\underline{X}(\vk ) = 
\left(
\begin{array}{cccc}
-c_1\vert \Psi_{\vk}\vert  & c_1\Psi_{\vk }^* & 0 & 0 \\
c_2\vert \Psi_{\vk }\vert  & c_2 \Psi_{\vk }^* & 0 & 0 \\
0 & 0 & -c_3^*\vert \Psi_{\vk }\vert  & c_3^*\Psi_{\vk }^*\\
0 & 0 &c_4^*\vert \Psi_{\vk }\vert  & c_4^* \Psi_{\vk }^*  \\
\end{array} \right)
\end{equation}
and
\begin{equation}
\underline{X}^{-1}(\vk ) =
\frac{1}{2\Psi_{\vk }^*}\left(
\begin{array}{cccc}
-\psi_{\vk}^* /c_1  & \psi_{\vk}^*/c_2 & 0 & 0 \\
1/c_1  & 1/c_2 & 0 & 0 \\
0 & 0 & -\psi_{\vk}^*/c_3^*  & \psi_{\vk}^*/c_4^*\\
0 & 0 &1/c_3^*  & 1/c_4^*  \\
\end{array} \right),
\end{equation}
where $\psi_{\vk }=\Psi_{\vk }/\vert \Psi_{\vk }\vert$.  
With $M=2$, the symmetry relations \cite{fishmanbook18} $X^{-1}(\vk )^*_{rn}=X^{-1}(-\vk )_{r+M,n+M}$ require that 
$c_3=c_1$ and $c_4=c_2$.  In addition,
\begin{equation}
\underline{X}(\vk )\cdot \underline{N}\cdot \underline{X}^{\dagger }(\vk ) =\underline{N}
\label{xin1}
\end{equation}
requires that $\vert c_1\vert^2 = \vert c_2\vert^2 = 1/2\vert \Psi_{\vk }\vert^2$, which produces Eq.\,(8)
in the main paper.

Notice that Eq.\,(S24) can be rewritten as 
\begin{equation}
\underline{X}^{-1}(\vk )\cdot \underline{N}\cdot \underline{X}^{-1\,\dagger }(\vk ) =\underline{N},
\label{xin2}
\end{equation}
which leads to
\begin{equation}
\sum_{r=1}^M \Bigl\{ \vert X^{-1}(\vk )_{rn}\vert^2 - \vert X^{-1}(\vk )_{r+M,n}\vert^2\Bigr\}=1.
\label{sumx}
\end{equation}
This expression allows us to rewrite the general result for the OAM of mode $n$ as
\begin{eqnarray}
{\cal L}_{zn}&&(\vk )=\frac{\hbar}{2}\sum_{r=1}^M\Bigl\{ X^{-1}(\vk )_{rn}\, \lk \, X^{-1}(\vk )^*_{rn}\nonumber \\
&&-X^{-1}(\vk)_{r+M,n}\,\lk \,X^{-1}(\vk )^*_{r+M,n}\Bigr\}\nonumber \\
&&\,\,\,\,\,\,\,\,\, = -\frac{i\hbar }{2} \sum_{r=1}^M \Bigl\{ \re X^{-1}(\vk )_{rn}\, \lk \,\im X^{-1}(\vk )_{rn}\nonumber \\
&&-\im X^{-1}(\vk )_{rn}\, \lk \,\re X^{-1}(\vk )_{rn }\nonumber \\ 
&&-\re X^{-1}(\vk )_{r+M,n}\, \lk \,\im X^{-1}(\vk )_{r+M,n}\nonumber \\
&&+\im X^{-1}(\vk )_{r+M,n}\, \lk \,\re X^{-1}(\vk )_{r+M,n }\Bigr\}.
\end{eqnarray}
So ${\cal L}_{zn}(\vk )$ vanishes if both matrix elements $X^{-1}(\vk )_{rn}$ and 
$X^{-1}(\vk )_{r+M,n}$ are real.

Returning to the FM zig-zag model (case $i$), $\Psi_{\vk }$ is complex except when $J_1=J_2$ and 
$\Psi_{\vk }=(\cos(k_xa)+\cos(k_ya))/2$.  Since the matrix elements
$X^{-1}(\vk )_{rn}$ and $X^{-1}(\vk )_{r+M,n}$ are then also real,
magnons of the square-lattice FM carry no OAM.   Similar conclusions follow for magnons of the square-lattice AF.

\section*{Analytic results for the Spin-Spin Correlation function of the AF Honeycomb Lattice}

For the AF honeycomb lattice (case $iv$), it is straightforward to evaluate the spin-spin correlation function using the method in Ref.\,\cite{fishmanbook18}.
As expected, only the transverse $xx$ and $yy$ matrix elements contribute to $S_{\alpha \beta }(\vk ,\omega )$ evaluated at the 
degenerate magnon frequency:
\begin{equation}
S_{\alpha \beta }(\vk ,\omega )=\delta_{\alpha \beta }
\sum_n S_{\alpha \alpha }^{(n)}(\vk ) \, \delta (\omega -\omega_n(\vk ))
\end{equation}
where $S_{zz}^{(n)}(\vk )=0$ and 
\begin{eqnarray}
S_{xx}^{(1)}(\vk )&=&S_{yy}^{(1)}(\vk ) \nonumber \\
&=&  \frac{S}{4}\Bigl\{ \vert X^{-1}(\vk )_{11}\vert^2 + \vert X^{-1}(\vk )_{41}\vert^2 \Bigr\}\nonumber \\
&=&\frac{S}{4f_{\vk }}(1+\kappa ),
\end{eqnarray}
\begin{eqnarray}
S_{xx}^{(2)}(\vk )&=&S_{yy}^{(2)}(\vk ) \nonumber \\
&=& \frac{S}{4}\Bigl\{ \vert X^{-1}(\vk )_{22}\vert^2 + \vert X^{-1}(\vk )_{32}\vert^2 \Bigr\}\nonumber \\
&=& \frac{S}{4f_{\vk }}(1+\kappa ),
\end{eqnarray}
with $f_{\vk }=\sqrt{(1+\kappa )^2-\vert \Gamma_{\vk }\vert^2 }$.  So the spin-spin correlation functions for the major and minor 
bands are the same.

\begin{figure}
\begin{center}
\includegraphics[width=8cm]{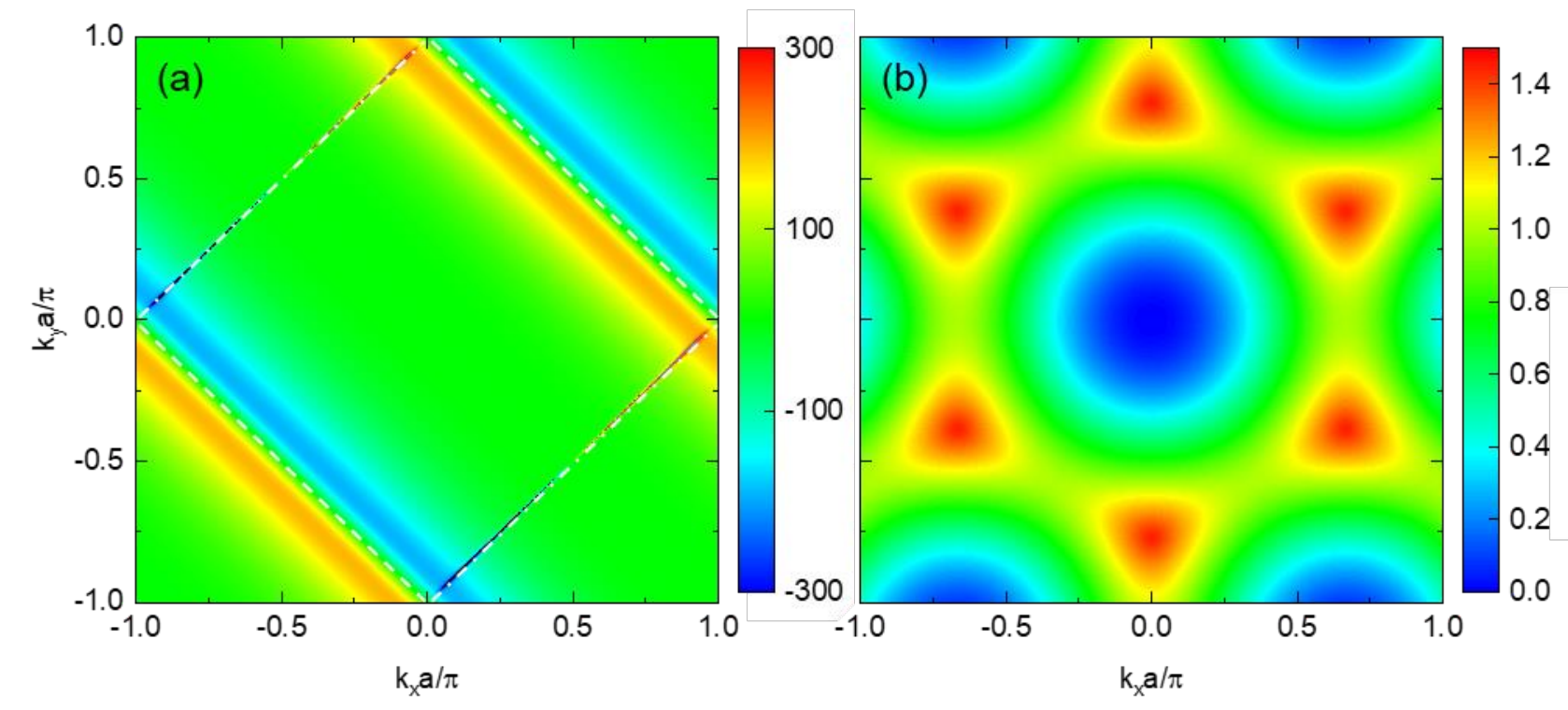}
\end{center}
\caption{Berry curvature of the lower magnon band for 
(a) a FM zig-zag lattice (case $i$) with $J_2/J_1=1.5$, and (b) a FM honeycomb lattice (case $iii$) with $d=-2D/3J=0.0067$. 
In panel (a), broken lines and dash-dot lines, respectively, indicate $k_x-k_y=\pm \pi/a$, where the magnon gap is 
opened by $J_2/J_1 \ne 1$, and $k_x+k_y=\pm \pi/a$, where the magnon gap is always closed.
The Berry curvature diverges along the $k_x+k_y=\pm \pi/a$ lines. }
\label{FigBerry}
\end{figure}

\section*{Comparison with Berry Curvature}

The Berry curvature of a multiband system in momentum space \cite{Xiao10} is given by
\begin{eqnarray}
\Omega_{n\vk}&=&\frac{i}{2 \pi} 
\sum_{m (\ne n)} \biggl\{ \frac{\langle n| \hat v_{x {\rm \bf k}}|m \rangle \langle m| \hat v_{y {\rm \bf k}} |n \rangle 
}{(\varepsilon_{m {\rm \bf k}} - \varepsilon_{n {\rm \bf k}})^2} \nonumber \\
&-& \frac{\langle n| \hat v_{y {\rm \bf k}}|m \rangle \langle m| \hat v_{x {\rm \bf k}} |n \rangle 
}{(\varepsilon_{m {\rm \bf k}} - \varepsilon_{n {\rm \bf k}})^2} \biggr\}, 
\label{EqBerry}
\end{eqnarray}
where, $n$ and $m$ are band indices, $\varepsilon_{n {\rm \bf k}}$ is the dispersion of band $n$, and 
$\hat v_{\eta {\rm \bf k}}$ is the velocity operator.
With the Hamiltonian written as $H=\sum_{\vk }{\hat H}_{\vk }$, $\hat v_{\eta {\rm \bf k}} = \partial \hat H_{\rm \bf k}/\partial k_\eta$. 

In order to highlight the difference between the OAM and the Berry curvature, we consider the  
FM zig-zag and honeycomb lattice models. 
For the FM zig-zag lattice model (case $i$) with $J_2/J_1=1.5$, the Berry curvature in Fig.\,S1(a) diverges where the gap closes between two bands at $k_x - k_y = \pm \pi/a$. 
To suppress this divergence, a small quantity $\sim$10$^{-4}$ is introduced in the denominators of the gauge-invariant form of $\Omega_{n\vk }$ above. 
Except for this divergence, the Berry curvature is quite flat, with some intensity modulations where the magnon gap is opened by $J_2/J_1 \ne 1$ at $k_x + k_y = \pm \pi/a$.
The Berry curvature changes sign at $k_x + k_y = \pm \pi/a$. 
As shown in the main text, the magnon OAM for the FM zig-zag lattice with $J_2/J_1 \ne 1$ has peak intensity at $k_x + k_y = \pm \pi/a$ and changes sign across this line at $k_x=k_y$. 

For the FM honeycomb lattice (case $iii$) with $d=-2D/3J = 0.0067$, 
the Berry curvature in Fig.\,S1(b) has sixfold symmetry and peaks at the K points. 
On the other hand, the magnon OAM has threefold symmetry, that is the K and K$^{\prime }$ points are different. 
Thus, while both the magnon OAM and Berry curvature exhibit some topological nature originating from the Berry connection, 
they capture different aspects of the system topology.

\section*{Periodic functions $\kov_x$ and $\kov_y$}

\begin{figure}
\begin{center}
\includegraphics[width=8cm]{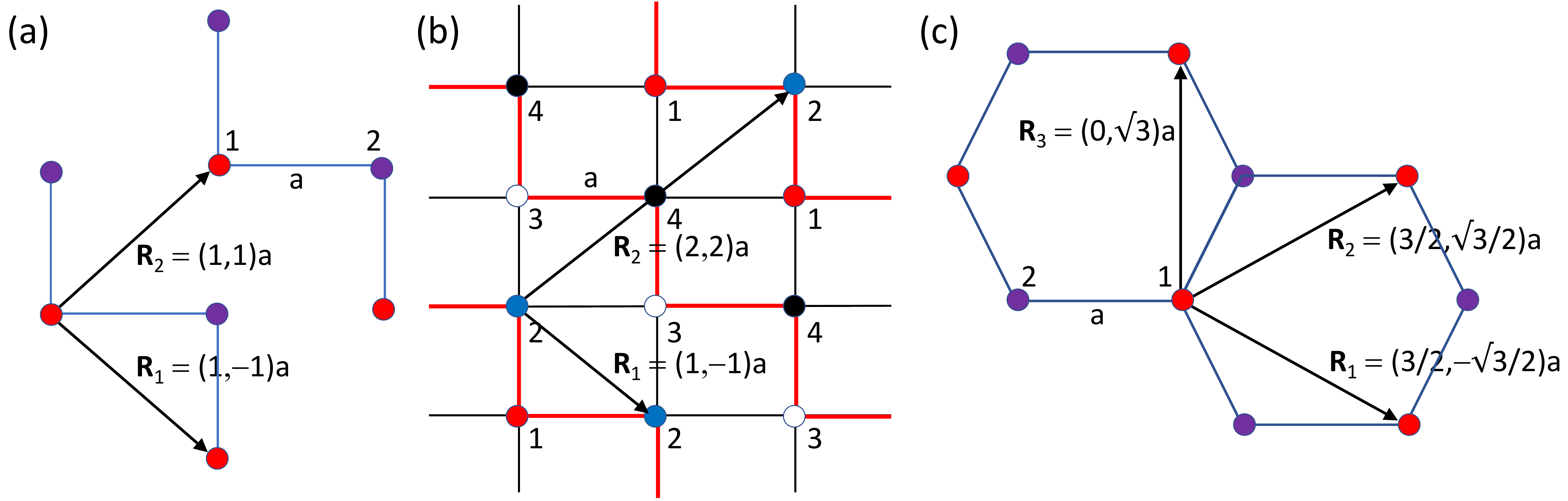}
\end{center}
\caption{(a) A FM zig-zag lattice with alternating exchange interactions, two distinct lattice sites, and
the two translation vectors ${\bf R}_1 = (1,-1)a$ and ${\bf R}_2 = (1,1)a$.  (b)
An AF zig-zag lattice (spins 1 and 2 up, spins 3 and 4 down) with alternating exchange interactions, four distinct lattice sites, and the two translation vectors 
${\bf R}_1 = (1,-1)a$ and ${\bf R}_2 = (2,2)a$.
(c) A honeycomb lattice with
the three translation vectors ${\bf R}_1 = (3/2,-\sqrt{3}/2)a$, ${\bf R}_2 = (3/2,\sqrt{3}/2)a$,
and ${\bf R}_3 = (0,\sqrt{3})a$.  In all three cases, translation vectors couple a specified site to neighboring sites of the same type.}
\label{Figdiscrete}
\end{figure}

On a discrete lattice, the continuous derivative $\partial /\partial x_\alpha$ should be replaced by a finite difference.
Fig.\,S2 sketches the distinct lattice translation vectors ${\bf R}_l$ that couple site 1 to other sites of type 1 for 
for the zig-zag and honeycomb lattices.
The finite difference of a discrete function $f(\vx)$ produced by translation vector ${\bf R}_l$ is then
\begin{eqnarray}
\delta_l f(\vx) = \frac{1}{2 |{\bf R}_l|} \Bigl\{ f(\vx + {\bf R}_l) - f(\vx - {\bf R}_l) \Bigr\}.
\end{eqnarray}
The continuous derivative $\partial f(\vx )/\partial x_{\alpha }$ along the $\alpha$ direction
is converted into a finite difference by the summation
\begin{eqnarray}
\Delta_\alpha f(\vx) &=& \sum_l \frac{R_{l\alpha}}{\vert R_l\vert }\,\delta_l f(\vx ) \\
&=& \sum_l
\frac{R_{l \alpha}}{2 |{\bf R}_l|^2} \Bigl\{ f(\vx + {\bf R}_l)
- f(\vx - {\bf R}_l) \Bigr\}, \nonumber 
\end{eqnarray}
where $R_{l \alpha}$ is the projection of the lattice translation vector ${\bf R}_l$ along the $\alpha$ axis.
Note that the finite difference $\Delta_\alpha f(\vx)$ approaches the continuous derivative $\partial f(\vx )/\partial x_{\alpha }$
when the lattice translation vectors are orthogonal and their size vanishes.
 
The finite difference of the factor $\exp ( i \vk \cdot \vx) $ that enters 
the Fourier transform of a magnon annihilation operator is given by
\begin{eqnarray}
\Delta_\alpha \exp ( i \vk \cdot \vx) &=& i \exp ( i \vk \cdot \vx) 
\sum_l \frac{R_{l \alpha}}{|{\bf R}_l|^2} \sin (\vk \cdot {\bf R}_l)\nonumber \\
&=& i\kov_{\alpha }\exp ( i \vk \cdot \vx),
\end{eqnarray}
where 
\begin{equation}
\kov_{\alpha } = \sum_l \frac{R_{l\alpha}}{|{\bf R}_l|^2} \sin (\vk \cdot {\bf R}_l).
\end{equation}
A periodic expression for ${\cal L}_{zn}(\vk )$ is obtained by using the revised OAM operator
\begin{equation}
\lk = -i \biggl( \kov_x\frac{\partial }{\partial k_y}-\kov_y\frac{\partial}{\partial k_x}\biggr).
\end{equation}
In the limit $N\rightarrow \infty $, the {\it momentum-space} derivatives
$\partial/\partial k_x$ and $\partial /\partial k_y$ do not need to be replaced by their finite differences.

\begin{figure}
\begin{center}
\includegraphics[width=8cm]{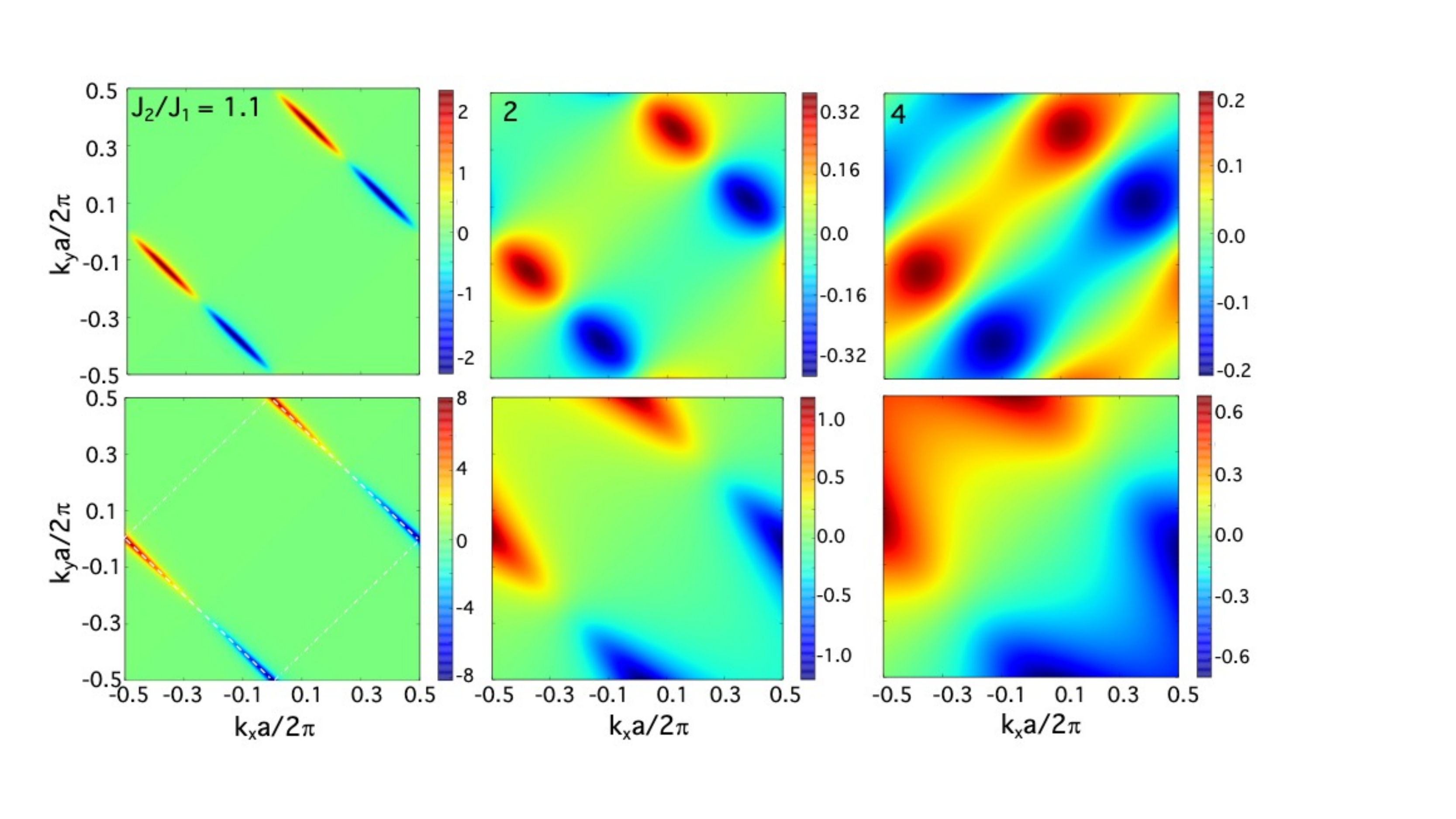}
\end{center}
\caption{A comparison between the OAM for the FM zig-zag lattice (case $i$) with different values of $J_2/J_1$ using periodic (top) and non-periodic (bottom) expressions for
$k_x$ and $k_y$. 
In the lower-left panel, broken lines and dash-dot lines, respectively, indicate $k_x-k_y=\pm \pi/a$, where the magnon gap is open by $J_2/J_1 \ne 1$, 
and $k_x+k_y=\pm \pi/a$, where the magnon gap is always closed.}
\label{FigCompare}
\end{figure}

Figure S3 plots the OAM of the FM zig-zag model (case $i$) for three values of $J_2/J_1$.  Summing over the two
translation vectors in Fig.\,S2(a), we obtain
\begin{eqnarray}
\kov_xa&=&\sin(k_xa) \cos(k_ya),\nonumber\\
\kov_ya&=&\sin(k_ya) \cos(k_xa).
\end{eqnarray}
To construct the top three panels, the non-periodic $k_x$ and $k_y$ have been replaced by $\kov_x$ and $\kov_y$ in the OAM operator $\lk $.
When non-periodic functions $k_x$ and $k_y$ are retained in the OAM operator, 
the OAM plotted in the bottom three panels of Fig.\,S3 is not bounded and has a much
larger magnitude than in the top three panels.  Hence, the periodic functions $\kov_x$ and
$\kov_y$ impose a bound on the OAM.

For the AF zig-zag lattice with four distinct lattice sites, summing over two translation 
vectors in Fig.\,S2(b) gives the periodic wavevectors
\begin{eqnarray}
\kov_xa&=&\frac{1}{2}\sin(k_xa-k_ya)\nonumber \\
&+&\frac{1}{4}\sin (2k_xa+2k_ya),\nonumber\\
\kov_ya&=&-\frac{1}{2}\sin(k_xa-k_ya)\nonumber\\
&+&
\frac{1}{4}\sin(2k_xa+2k_ya ).
\end{eqnarray}
For the FM or AF honeycomb lattice, summing over the three translation vectors in Fig.\,S2(b) gives
\begin{eqnarray}
\kov_xa&=&\sin(3k_xa/2)\cos(\sqrt{3}k_ya/2),\nonumber \\
\kov_ya&=& \frac{1}{\sqrt{3}}\Bigl\{ \sin(\sqrt{3}k_ya/2)\cos(3k_xa/2) \nonumber \\
&&+\sin(\sqrt{3}k_ya)\Bigr\}.
\end{eqnarray}
In the limit of small $k_x$ and $k_y$, $\kov_x\rightarrow k_x$ and $\kov_y\rightarrow k_y$ for the
zig-zag lattices while $\kov_x\rightarrow 3k_x/2$ and $\kov_y\rightarrow 3k_y/2$ for the honeycomb
lattices.


\begin{thebibliography}{10}
\bibitem{pseudo}
Strictly speaking, the orbital angular momentum of a crystal discussed in this
  work is the ``pseudo" orbital angular momentum \cite{Streib21} because
  rotational invariance is violated in a lattice.
\end{thebibliography}
\end{document}